\documentclass[12pt,twoside]{article}
\usepackage{xrb2007}
\usepackage{epsf}
\begin{document}

\session{Jets}

\shortauthor{Rothstein and Lovelace}
\shorttitle{Assembling the Ingredients for a Jet}

\title{Assembling the Ingredients for a Jet: How Do Large-Scale Magnetic Fields Get There and What Happens When They Do?}
\author{David M. Rothstein}
\affil{NSF Astronomy and Astrophysics Postdoctoral Fellow, Department of Astronomy, Cornell University, Ithaca, NY 14853-6801; droth@astro.cornell.edu}
\author{Richard V. E. Lovelace}
\affil{Departments of Astronomy and Applied and Engineering Physics, Cornell University, Ithaca, NY 14853-6801; RVL1@cornell.edu}

\begin{abstract}
Jets and outflows are observed around a wide variety of accreting objects and seem to be a near-ubiquitous feature of accretion disks.  Large-scale magnetic fields are thought to be necessary for jet formation in many systems, but a longstanding puzzle is that the turbulence which is responsible for inward disk accretion should be even more efficient at causing a large-scale magnetic field to diffuse outward; it just doesn't seem possible to build up a strong field in the inner disk through advection of a weak one from outside.  Here, we report theoretical work which challenges this conventional wisdom and shows that the surface layer of the disk (which is nonturbulent) can easily advect magnetic fields inward.  This opens up the possibility for more diverse and robust jet formation than is seen in numerical simulations which assume that the large-scale field must be generated entirely via a dynamo within the inner disk.  In addition, strong disk turbulence (as inferred from observations) can naturally be explained by the magnetorotational instability feeding off a large-scale magnetic field anchored in the surface layer.
\end{abstract}

\section{Introduction}
The origin of jets and outflows is a fundamental problem in astrophysics.  Theoretical models for collimated, relativistic jets, dating back over 30 years, usually require that the disk be threaded by a strong, large-scale magnetic field in the inner regions \citep[e.g.,][]{Lovelace1976,Blandford1976}.  In the early years, it was often assumed that this configuration could arise as a result of an initially weak magnetic field supplied in the outer disk and advecting inward along with the accretion flow.  In that case, simple geometric considerations show that the field will be compressed and strengthened as it approaches the inner disk.  Furthermore, a large-scale field threading the disk could be expected to drive centrifugal winds at all radii and assist in the process of angular momentum extraction \citep[e.g.,][]{BlandfordPayne1982}.

More recently, however, the idea that large-scale magnetic fields can be advected inward has fallen out of favor.  The essential problem is that accretion disks are almost certainly turbulent, which makes them very poor conductors.  The turbulent motions within the disk lead to enhanced reconnection of the large-scale magnetic field and a large turbulent diffusivity; the result is that any concentration of large-scale magnetic field will quickly diffuse away \citep*[e.g.,][]{vanball1989,Lubow1994}.  This problem can be avoided if the field is strong enough \citep*{Lovelace1994} or if it is highly nonaxisymmetric \citep{Spruit2005}, but it cannot be avoided if one wishes to start with an {\it initially} weak field in the outer regions of a thin, axisymmetric disk and allow it to be advected inward and compressed.  Thus, it has come to be believed that any large-scale magnetic field in the inner disk must arise there locally, which severely limits the possible ways in which jets and winds can form.

Numerical simulations do not do as good a job at addressing this puzzle as one might hope, for reasons which are primarily technical.  Most three-dimensional magnetohydrodynamic (MHD) simulations performed to date begin with a geometry in which the matter and magnetic field are confined to a torus far away from the outer boundary of the simulation.  Without any net magnetic flux in the initial geometry, and without any magnetic field supplied at the outer boundary, it is impossible to determine whether or not a large-scale magnetic field would be advected inward.  Nonetheless, in one simulation in which a weak, large-scale magnetic field {\it was} injected at the outer boundary, the field was clearly observed to be dragged inward and to build up to strong values in the inner disk \citep*{Igumenshchev2003}.  Furthermore, even in simulations which begin with a torus, the character of the field geometry in the initial torus has a significant influence on the jet structures that develop near the central black hole after the material in the torus has begun to advect inward \citep*{McKinney2004,Beckwith2007}.

In light of the current state of affairs, we are motivated to reconsider the problem of magnetic field advection in an accretion disk.  The results of our work are presented in \citet{Rothstein2008} and are summarized here.

\section{Magnetic Field Advection in the Surface Layer of an Accretion Disk}
The simple arguments outlined above, which conclude that turbulent diffusion will prevent inward advection of the magnetic field, do not take into account the disk's vertical structure.  In fact, this is a key omission.  Since magnetic fields are sustained by the flow of current, not mass, even rarefied regions far away from the main central body of the disk can affect the inward advection of magnetic fields if they have current flowing through them.

In this vein, \citet{OgilvieLivio2001} pointed out that the vertical profile of the diffusivity is a key quantity that must be computed in order to determine whether or not the magnetic field can advect inward.  \citet{BKL2007}, operating under the assumption that the surface layer of the accretion disk is nonturbulent and therefore highly conducting, suggested that the greatly reduced diffusivity at the surface could allow the magnetic field to advect inward there.  Our recent work \citep{Rothstein2008} examines this possibility more carefully; we consider the physically realistic case of an accretion disk driven by the magnetorotational instability \citep[MRI;][]{BalbusHawley1991} and show that under very general conditions, the base of the nonturbulent corona above such a disk should participate in the accretion flow and sustain the inward advection of a weak, large-scale magnetic field.

\begin{figure}[!ht]
\plotone{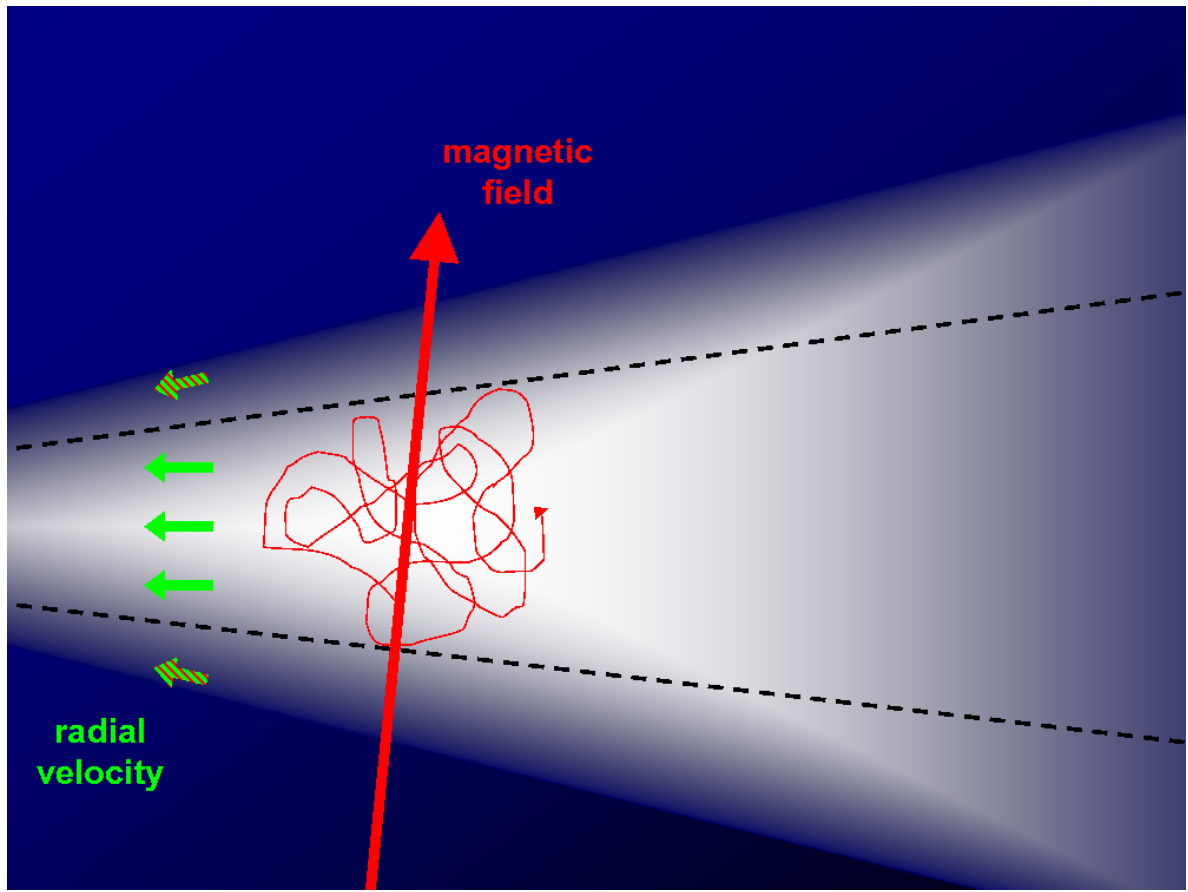}
\caption{Sketch of a weak magnetic field threading an accretion disk.  At a large enough height above the equatorial plane (indicated by the dashed lines), the mass density is low enough for the plasma to become magnetically dominated and for the magnetorotational turbulence to be suppressed.  The main body of the disk accretes inward due to turbulent angular momentum transport (solid arrows), while large-scale magnetic stresses remove angular momentum vertically from the base of the nonturbulent corona, causing it to advect inward as well (striped arrows) and carry the large-scale magnetic field with it.}
\label{fig:geometry}
\end{figure}

The essential physical ideas are illustrated in Figure \ref{fig:geometry}, which shows a weak magnetic field threading an accretion disk.  The main body of the disk is MRI turbulent and therefore highly diffusive.  However, the mass density in the disk drops off with height more rapidly than the magnetic field strength, and at a certain height above the equatorial plane (indicated by the dashed lines), the vertical magnetic energy density will become comparable to the local gas pressure.  Since the MRI is a weak field instability \citep{BalbusHawley1991}, the field in this region will be strong enough, relatively speaking, to suppress the turbulence.  Such magnetically-dominated coronas above a weakly-magnetized disk have now been observed in a wide variety of numerical simulations \citep*[e.g.,][]{MillerStone2000,DeVilliers2003,Hirose2004,McKinney2004,Fromang2006,Hirose2006,Krolik2007}.  

Clearly, if the base of this nonturbulent corona participates in the accretion flow, as indicated by the striped arrows in Figure \ref{fig:geometry}, it can carry the large-scale magnetic field inward without being opposed by turbulent diffusion.  We therefore must ask:  What determines the direction of the radial velocity in this region?  In the main body of the disk, an inward radial velocity is achieved via turbulent angular momentum transport, but this process is, by definition, not available in the corona.  However, the large-scale magnetic field {\it itself} acts as an efficient conduit for the vertical transport of angular momentum, both between the main body of the disk and the base of the corona and between the corona and the wind region above it.  Thus, the stress due to the large-scale field itself can drag the base of the nonturbulent region inward.

A key feature of this process is the fact that the MRI is a weak field instability; by definition, the field at the base of the nonturbulent region must be strong enough to affect the local gas dynamics, regardless of how weak it is compared to the thermal pressure in the main body of the disk.  Consulting Figure \ref{fig:geometry}, we see that if the large-scale field becomes weaker, more of the disk will become turbulent (i.e., the dashed lines will move higher); thus, this weaker field will only be ``responsible'' for advecting inward a smaller amount of material.  Whether or not advection occurs is thus primarily a question of geometry rather than magnetic field strength (but see \S \ref{sec:limits}).  Analysis of the equations of motion for a thin disk \citep{Rothstein2008} shows that, to an order of magnitude, the base of the nonturbulent corona will advect inward if $\left( T_{mag} \right)_{\phi z} / \left( P_{mag} \right)_{z} > v_{z} / v_{K}$, where $\left( T_{mag} \right)_{\phi z}$ is the azimuthal-vertical magnetic stress at the base of the corona, $\left( P_{mag} \right)_{z}$ is the pressure of the vertical magnetic field there, and $v_{z}/v_{K}$ is the ratio of the vertical velocity at the base of the corona to the Keplerian orbital velocity.  (This latter term opposes inward advection; it arises because material from closer to the equatorial plane rising upward through the corona carries excess angular momentum that tends to fling it centrifugally outward.)

Comparison with vertically-stratified shearing box simulations of \citet{MillerStone2000} suggests that the above condition will be very easy to satisfy even without an initial imposed large-scale vertical field.  Thus, the magnetic field can be advected inward.  However, an important caveat is that the vertical angular momentum transport must be in the correct direction in order for the base of the corona to be dragged inward.  Although an imposed ``dipole-type'' field (i.e., where the field lines penetrate through the disk, as in Figure \ref{fig:geometry}) will lead to vertical magnetic stress of the correct sign, other magnetic field geometries (such as a quadrupole-type geometry) may not be conducive to inward advection.

\section{Limits on the Magnetic Field Advection} \label{sec:limits}

The above discussion assumes that the material at the base of the nonturbulent corona is moving in Keplerian orbits, as in standard thin disk theory.  If the large-scale magnetic field is extremely weak, however, the base of the nonturbulent region may be so high above the equatorial plane, and the plasma there may be so rarefied, that orbits are no longer circular; in this case, radiation pressure can prevent the large-scale field from advecting inward.  As shown in \citet{Rothstein2008}, for a fiducial distance of $\sim 10^{2}$ Schwarzschild radii from a $10^{8} M_{\sun}$ black hole (or $\sim 10^{5}$ Schwarzschild radii from a $10 M_{\sun}$ star) the vertical magnetic field at the disk surface must be $\gg 0.1$ G in order to overcome radiation pressure and advect inward.  Note that this is many orders of magnitude ($\sim 10^5$) smaller than the equipartition field strength in the main body of the disk, so magnetic buoyancy acting on the dynamo-generated disk field may play a role in meeting this condition.

Another limit comes from non-ideal MHD.  Although Ohmic diffusion in a fully ionized corona is negligible, ambipolar diffusion may become important in the corona of a partially ionized disk.  This imposes a typical limit of $\left( B_{z} / {\rm G} \right)^{2} f_{i} \gg 10^{-3} m_{8}^{-1} r_{3}^{-5/2}$, where $B_{z}$ is the vertical field and $f_{i}$ is the ionization fraction at the base of the corona, $m_{8}$ is the mass of the central object in $10^{8}$ solar masses, and $r_{3}$ is the distance from the central object in units of $10^{3}$ Schwarzschild radii \citep{Rothstein2008}.

\section{Conclusions}

Using simple dynamical arguments and well-known properties of the magnetorotational instability, we are led to the conclusion that a weak, large-scale magnetic field threading a thin, turbulent accretion disk can easily be advected inward in the nonturbulent surface layer.  This has several interesting consequences.

It is well known that a magnetic field threading an accretion disk can drive strong magnetorotational turbulence \citep*{Hawley1995}.  In fact, recent analyses suggest that without an externally-imposed vertical field, the turbulence is much weaker than that which is inferred from observations \citep*{King2007,Pessah2007,Fromang2007}, which would require that a vertical field of a few percent of equipartition threads the disk.  Such vertical fields could naturally be explained by advection, however.  If an advected field is sustained by currents flowing in the nonturbulent surface layer of the disk, as we have argued, then the main, turbulent body of the disk would see it as a fixed, ``externally-imposed'' seed field for the MRI.  Furthermore, we would then expect the (turbulent) variability timescales of the disk to evolve in response to the history of magnetic field advection; this may be related to the processes discussed by \citet{Tagger2004} to explain the complex variability of bright X-ray binaries such as GRS~1915+105.

Perhaps most fundamentally, the ease with which large-scale magnetic fields can advect inward in an accretion disk could help explain the origin of jets and winds from a wide variety of accreting objects, including black holes, neutron stars, white dwarfs and young stars.  Numerical simulations of jet formation currently focus on black holes, and the jets that arise are confined largely to the inner region, where the specific properties of the black hole have a large influence.  If an advected magnetic field is introduced into the picture, perhaps this may help explain the near-ubiquity of jets and outflows from a wide variety of accretion disks, regardless of the central object.

\acknowledgements D.~M.~R. would like to thank the organizers of the conference for a unique and refreshing meeting.  We also thank  G.~S. Bisnovatyi-Kogan, M.~M. Romanova, and I.~G. Igumenshchev for valuable discussions.  D.~M.~R. is supported by an NSF Astronomy and Astrophysics Postdoctoral Fellowship under award AST-0602259.  R.~L. is supported in part by NASA grants NAG5-13220 and NAG5-13060 and by NSF grant AST-0507760.


\begin{thebibliography}{}
\bibitem[Balbus \& Hawley(1991)]{BalbusHawley1991} Balbus, S.~A., \& Hawley, J.~F.\ 1991, \apj, 376, 214 
\bibitem[Beckwith et al.(2007)Beckwith, Hawley, \& Krolik]{Beckwith2007} Beckwith, K., Hawley, J.~F., \& Krolik, J.~H.\ 2007, \apj, submitted (arXiv:0709.3833)
\bibitem[Bisnovatyi-Kogan \& Lovelace(2007)]{BKL2007} Bisnovatyi-Kogan, G.~S., \& Lovelace, R.~V.~E.\ 2007, \apjl, 667, L167 
\bibitem[Blandford(1976)]{Blandford1976} Blandford, R.~D.\ 1976, \mnras, 176, 465 
\bibitem[Blandford \& Payne(1982)]{BlandfordPayne1982} Blandford, R.~D., \& Payne, D.~G.\ 1982, \mnras, 199, 883
\bibitem[De Villiers et al.(2003)De Villiers, Hawley, \& Krolik]{DeVilliers2003} De Villiers, J.-P., Hawley, J.~F., \& Krolik, J.~H. 2003, \apj, 599, 1238
\bibitem[Fromang \& Nelson(2006)]{Fromang2006} Fromang, S., \& Nelson, R.~P.\ 2006, \aap, 457, 343 
\bibitem[Fromang \& Papaloizou(2007)]{Fromang2007} Fromang, S., \& Papaloizou, J.\ 2007, \aap, 476, 1113 
\bibitem[Hawley et al.(1995)Hawley, Gammie, \& Balbus]{Hawley1995} Hawley, J.~F., Gammie, C.~F., \& Balbus, S.~A.\ 1995, \apj, 440, 742 
\bibitem[Hirose et al.(2004)]{Hirose2004} Hirose, S., Krolik, J.~H., De Villiers, J.-P., \& Hawley, J.~F. 2004, \apj, 606, 1083
\bibitem[Hirose et al.(2006)Hirose, Krolik, \& Stone]{Hirose2006} Hirose, S., Krolik, J.~H., \& Stone, J.~M.\ 2006, \apj, 640, 901 
\bibitem[Igumenshchev et al.(2003)Igumenshchev, Narayan, \& Abramowicz]{Igumenshchev2003} Igumenshchev, I.~G., Narayan, R., \& Abramowicz, M.~A. 2003, \apj, 592, 1042
\bibitem[King et al.(2007)King, Pringle, \& Livio]{King2007} King, A.~R., Pringle, J.~E., \& Livio, M.\ 2007, \mnras, 376, 1740
\bibitem[Krolik et al.(2007)Krolik, Hirose, \& Blaes]{Krolik2007} Krolik, J.~H., Hirose, S., \& Blaes, O.\ 2007, \apj, 664, 1045 
\bibitem[Lovelace(1976)]{Lovelace1976} Lovelace, R.~V.~E. 1976, Nature, 262, 649
\bibitem[Lovelace et al.(1994)Lovelace, Romanova, \& Newman]{Lovelace1994} Lovelace, R.~V.~E., Romanova, M.~M., \& Newman, W.~I. 1994, \apj, 437, 136
\bibitem[Lubow et al.(1994)Lubow, Papaloizou, \& Pringle]{Lubow1994} Lubow, S.~H., Papaloizou, J.~C.~B., \& Pringle, J.~E. 1994, \mnras, 267, 235
\bibitem[McKinney \& Gammie(2004)]{McKinney2004} McKinney, J.~C., \& Gammie, C.~F.\ 2004, \apj, 611, 977 
\bibitem[Miller \& Stone(2000)]{MillerStone2000} Miller, K.~A., \& Stone, J.~M.\ 2000, \apj, 534, 398 
\bibitem[Ogilvie \& Livio(2001)]{OgilvieLivio2001} Ogilvie, G.~I., \& Livio, M.\ 2001, \apj, 553, 158 
\bibitem[Pessah et al.(2007)Pessah, Chan, \& Psaltis]{Pessah2007} Pessah, M.~E., Chan, C.-k., \& Psaltis, D.\ 2007, \apjl, 668, L51 
\bibitem[Rothstein \& Lovelace(2008)]{Rothstein2008} Rothstein, D.~M., \& Lovelace, R.~V.~E. 2008, \apj, in press (arXiv:0801.2158)
\bibitem[Spruit \& Uzdensky(2005)]{Spruit2005} Spruit, H.~C., \& Uzdensky, D.~A. 2005, \apj, 629, 960
\bibitem[Tagger et al.(2004)]{Tagger2004} Tagger, M., Varni{\`e}re, P., Rodriguez, J., \& Pellat, R.\ 2004, \apj, 607, 410 
\bibitem[van Ballegooijen(1989)]{vanball1989} van Ballegooijen, A.~A.\ 1989, in ASSL Vol.~156: Accretion Disks and Magnetic Fields in Astrophysics, ed. G. Belvedere (Dordrecht: Kluwer Academic Publishers), 99 
\end{thebibliography}
\end{document}